\documentclass[pra,preprint,showpacs,showkeys,nofootinbib,superscriptaddress,tightenlines]{revtex4}

\usepackage{ulem}
\usepackage{graphicx}  %
\usepackage{amsmath}
\usepackage{bm}  %

\newcommand{\bea}{\begin{eqnarray}}
\newcommand{\eea}{\end{eqnarray}}
\newcommand{\beq}{\begin{equation}}
\newcommand{\eeq}{\end{equation}}
\newcommand{\bqa}{\begin{eqnarray}}
\newcommand{\eqa}{\end{eqnarray}}

\def\mqo2{{\!\!\!}}

\begin{document}

\preprint{HISKP-TH-10-14}
\preprint{INT-PUB-10-024}
\title{Efimov Physics in Atom-Dimer Scattering of $^6$Li Atoms}

\author{H.-W. Hammer}
\affiliation{Helmholtz-Institut f\"ur Strahlen- und Kernphysik
        (Theorie) and Bethe Center for Theoretical Physics,
        Universit\"at Bonn, 53115 Bonn, Germany}
\affiliation{Institute for Nuclear Theory,
University of Washington, Seattle WA\ 98195 USA\\}
\author{Daekyoung Kang}
\affiliation{Department of Physics,
         The Ohio State University, Columbus, OH\ 43210, USA}
\affiliation{Institute for Nuclear Theory,
University of Washington, Seattle WA\ 98195 USA\\}

\author{Lucas Platter}
\affiliation{Institute for Nuclear Theory,
University of Washington, Seattle WA\ 98195 USA\\}

\date{\today}

\begin{abstract}
  $^6$Li atoms in the three lowest hyperfine states display universal
  properties when the S-wave scattering length between each pair of
  states is large. Recent experiments reported four pronounced
  features arising from Efimov physics in the atom-dimer relaxation rate,
  namely two resonances and two local minima.  We use the universal 
  effective field theory to calculate the atom-dimer relaxation rate
  at zero temperature. Our results describe
  the four features qualitatively and imply there is a hidden local
  minimum. In the vicinity of the resonance at 
  685 G, we perform a finite temperature calculation which 
  improves the agreement of theory and experiment. We
  conclude that finite temperature effects cannot be neglected 
  in the analysis of the experimental data.
\end{abstract}

\smallskip
\pacs{31.15.-p,34.50.-s, 67.85.Lm, 03.75.Ss}
\keywords{
Degenerate Fermi gases, three-body recombination,
scattering of atoms and molecules. }
\maketitle
\section{Introduction}\label{sec:intro}
Few-body systems with large two-body scattering length $a$ are of great
interest because they have universal properties insensitive to 
the details of the interaction at short distances. 
The simplest example is a universal two-body bound state with binding
energy $\hbar^2/(m a^2)$ when $a>0$. The corrections to this formula
are suppressed by $\ell/a$ where $\ell$ is the natural low-energy
length scale. For alkali atoms, $\ell$ is given by the van der Waals
length $\ell_{vdW}$ which quantifies the range of the interaction.
Examples of universal two-body states can be found
in different branches of physics such as atomic
physics, nuclear physics, and high-energy physics
\cite{Hammer:2010kp,Platter:2009gz}. Universal properties also exist 
for more than two particles. Systems of three
identical bosons have an infinite number of geometrically spaced 
low-energy bound states ({\it Efimov trimers}) with an accumulation 
point at zero energy when the scattering length is taken to infinity and 
the range of the interaction is taken to zero \cite{Efimov}.
In this so-called {\it unitary limit}, the binding energies of two
successive trimers are related by a multiplicative factor of
approximately $22.7^2\approx 515$. This remarkable feature is the
consequence of a discrete scaling symmetry with a scaling factor
$e^{\pi/s_0}\approx22.7$ where $s_0\approx 1.00624$ for identical
bosons. For more complicated systems, the scaling factor depends on the 
mass ratios and the spin of the components.
We will refer to phenomena that result from the
discrete scale invariance in the three-body sector as {\it Efimov
physics} \cite{Braaten:2004rn}.
Corrections to Efimov physics in a process with typical wave number 
$k$ are suppressed by $\ell/a$ and by $ka$.

Efimov trimers in ultracold atomic gases can be observed through
enhanced loss rates in three-body recombination
and atom-dimer relaxation processes \cite{NM-99,EGB-99,BBH-00}.
The first evidence for an Efimov trimer in a
system of bosonic $^{133}$Cs atoms was presented by Kraemer {\it et al.}\
\cite{Kraemer:2006}. In a subsequent experiment with a mixture of $^{133}$Cs
atoms and dimers, Knoop {\it et al.}\ observed a resonance in the loss of
atoms and dimers \cite{Knoop:2008} which can be explained by an Efimov trimer
crossing the atom-dimer threshold \cite{Helfrich:2009uy}. Efimov trimers have
also been observed using other types of bosonic atoms.  Zaccanti {\it et al}.\
\cite{Zaccanti:2008} found evidence for two adjacent Efimov trimers in a gas
of ultracold $^{39}$K atoms and Gross {\it et al.}\ \cite{Gross:2009} measured
two Efimov features across a Feshbach resonance and confirmed the universal
relation between the features with $^7$Li atoms.  Attached to each Efimov
trimer are two universal tetramers which can be observed in four-body loss
processes \cite{Platter:2004qn,Hammer:2006ct,Stecher:2008}.  
Ferlaino {\it et al.}\
\cite{Ferlaino:2009} observed signatures of two such tetramers close to an
Efimov trimer with $^{133}$Cs atoms. In $^7$Li atoms, Pollack {\it et al.}\
\cite{Pollack:2009} have seen two sets of two tetramers that are close to the
corresponding Efimov trimers.

Efimov physics can also occur in fermionic systems with three or more spin
states \cite{Efimov}. Several features associated with Efimov trimers in
$^6$Li atoms have been reported recently. Two three-body recombination
features were observed by Ottenstein {\it et al.}\ \cite{Heidelberg} and
Huckans {\it et al.}\ \cite{PennState}. These features were analyzed
theoretically in Refs.~\cite{Braaten:2008wd,schmidt:2008fz,Naidon:2009} and
traced back to an Efimov trimer close to the three-atom threshold
\cite{schmidt:2008fz,Naidon:2009,Braaten:2009ey}. Another recombination
feature was discovered by Williams {\it et al.}\ \cite{Williams:2009zzd} and
identified as a second trimer close to the threshold. Braaten {\it et al.}\
\cite{Braaten:2009ey} analyzed the energy spectrum of Efimov trimers and
predicted the crossings of the trimers with the atom-dimer threshold. These
crossings imply two resonances in atom-dimer scattering. The
resonances were identified in recent experiments by Lompe {\it et al.}\
\cite{Lompe:2010} and Nakajima {\it et al.}\ \cite{Nakajima:2010}.
Furthermore, two local minima in the atom-dimer relaxation rate were 
reported in Ref.~\cite{Lompe:2010}.

In this paper, we study Efimov physics in atom-dimer scattering of $^6$Li
atoms in the three lowest hyperfine states and perform a 
complete zero-range calculation at zero temperature. 
The size of finite range corrections to our results is estimated. We
identify the two resonances and the two minima in the atom-dimer relaxation
rate.  Quantitatively, our zero temperature results show
deviations from the data.  
An approximate finite temperature calculation near
one of the resonances improves the description of the data
considerably and shows that finite temperature
effects account for about 25\% of the discrepancy. The
paper is organized as follows: In Sec.~\ref{sec:relaxation} and
Sec.~\ref{sec:STMeq} we explain our theoretical framework. The numerical
results at zero temperature are displayed and compared with the experimental
data in Sec.~\ref{sec:results}. In Sec.~\ref{sec:themal}, we carry out an
approximate finite temperature calculation near the resonance at $B=685$ G.
We summarize our results and conclude in Sec.~\ref{sec:summary}.

\section{Dimer relaxation}
\label{sec:relaxation}
In this section we discuss the losses of atoms and dimers through inelastic
scattering processes and present expressions for the relaxation rate
constants.

Before proceeding with our discussion of atom-dimer scattering, we explain our
notation and terminology. We label an atom in one of the three hyperfine
states of the $^6$Li atoms with an index $i$ where $i=1,2$ or $3$. The S-wave
scattering length between atoms in states $i$ and $j$ is denoted either as
$a_{ij}$ or as $a_k$ where $k\neq i\neq j$. Two atoms in the same state can
not scatter in an S-wave because of the Pauli principle. If the scattering
length $a_{ij}$ is positive and much larger than the van der Waals length
$\ell_{\rm vdW}\approx 65a_0$, the atoms $i$ and $j$ can form a dimer with
binding energy $\hbar^2/(m a_{ij}^2)$. We call this dimer the ({\it shallow})
$ij$-dimer or simply a shallow dimer. Shallow dimers have to be distinguished
from deep dimers with binding energy of order $\hbar^2/(m \ell_{\rm vdW}^2)$
or larger.

In a gas of atoms $i$ and $jk$-dimers, the atoms and dimers can undergo
inelastic collisions into atoms and deeply bound dimers with a binding energy
larger than that of the $jk$-dimer. This inelastic process is called {\it
  dimer relaxation}. The difference in the binding energies of
the initial and final state dimers is released as kinetic energy and the 
atom and dimer in the final state recoil from each other. If their 
kinetic energies are
larger than the trapping potential, they escape the trap. The loss rate for
the number density $n_i$ of atoms $i$ and number density $n_{jk}$ of
$jk$-dimers is
\begin{equation}
\frac{d\ }{dt} n_i =\frac{d\ }{dt} n_{jk} 
= - \beta_i \,n_i n_{jk},
\label{dndt}
\end{equation}
where the coefficient $\beta_i$ is the {\it relaxation rate constant}
for the $jk$-dimer and atom $i$.

In the case of identical bosons, dimer relaxation is possible
only if the final state consists of an atom and a deep dimer.  
However, in the three-fermion system relaxation can also occur into
into shallow dimers. For example, in a scattering process of an
atom $i$ and a $jk$-dimer, relaxation can proceed into the $ij$-dimer
provided the binding energy of the $ij$-dimer is larger than that of
the $jk$-dimer. Therefore, the total rate $\beta_i$ is the sum of all
relaxation rates into atoms plus shallow dimers and atoms plus deep
dimers
\begin{equation}
\beta_i=
\sum_{j\neq i} \beta^{\rm sh}_{i\to j}+ \beta^{\rm deep}_{i},
\label{beta:tot}
\end{equation}
where the index $i$ implies atom $i$ plus $jk$-dimer in the initial state
and the $j$ implies atom $j$ plus $ik$-dimer in the final state.  For
brevity, we refer to $\beta^{\rm sh}_{i\to j}$ as the rate into the shallow
dimer or the rate into the (final) $ik$-dimer and to $\beta^{\rm
  deep}_{i}$ as the rate into the deep dimers.
 
The relaxation rate can be calculated from the T-matrix element for atom-dimer
scattering ${\cal T}^{\rm AD}_{ij}(k,p;E)$ where $k$ and $p$ are the relative
wave numbers of the atom and dimer in initial and final state, respectively,
and $E$ is the total energy.  By using the {\it optical theorem}, we can
calculate the total rate for atom-dimer scattering which is the sum of elastic
and inelastic rates. 
However, in the low energy limit $k\to 0$, the elastic
rate scales as $k$ because ${\cal T}^{\rm AD}_{ii}(k,k;E)$ is constant
and the two-body phase space gives one power of $k$
\cite{Braaten:2004rn}.
The elastic rate therefore vanishes at zero energy and
the optical theorem gives the total relaxation rate $\beta_i$
\begin{equation}
\beta_i =\frac{2\hbar}{m}\,
{\rm Im} {\cal T}^{\rm AD}_{ii}(0,0,-1/(m a_i^2)).
\label{beta:opt}
\end{equation}
The rate into the shallow dimer $\beta^{\rm sh}_{i\to j}$ is determined by
the square of the T-matrix element multiplied by the two-body phase space:
\begin{equation}
\beta^{\rm sh}_{i\to j} =
\frac{2p\hbar}{3\pi m} 
\left| {\cal T}^{\rm AD}_{ij}(0,p,-1/(m a_i^2))
\right|^2
\,\theta(a_i-a_j),
\label{beta:sh}
\end{equation}
where the wave number $p=(2/\sqrt{3})\sqrt{a_j^{-2}-a_i^{-2}}$ and 
the $\theta$-function is inserted because the relaxation 
is allowed only when the initial dimer binding energy is smaller than the final
one. The prefactor $2p/(3\pi)$ comes from the two-body phase space integral.
From Eq.~(\ref{beta:sh}), one can deduce that $\beta^{\rm sh}_{i\to j}$
vanishes as $\sqrt{a_i-a_j}$ near the crossing of the $ik$- and $jk$-dimers
where $a_i$ approaches $a_j$.

The relaxation rate into deep dimers $\beta^{\rm deep}_i$ could also
be calculated by using the T-matrix in a similar way to
Eq.~\eqref{beta:sh} if the theory described deep dimers explicitly.
Alternatively, the effects of deep dimers can be taken into account
indirectly by using an analytic continuation of the three-body parameter
into the complex plane as introduced in
Ref.~\cite{Braaten:2002sr}. Then, the partial rate $\beta^{\rm
  deep}_i$ can be obtained by using the relation in Eq.~\eqref{beta:tot}.

D'Incao and Esry have calculated the scattering length dependence
of ultracold three-body collisions near overlapping Feshbach 
resonances for a variety of cases \cite{DIncao:2009}. For the 
relaxation rate into the shallow dimer $\beta^{\rm sh}_{i\to j}$, 
they find that it scales like $a_{j}^2/a_{i}$ times a log-periodic function
of $a_{j}$ if $a_{i} \gg a_{j}$ while $a_{k}$ is non-resonant.
The rate into deep dimers $\beta^{\rm deep}_i$ scales like $a_{j}^2/a_{i}$ 
times a prefactor that is a constant for positive $a_{j}$ and a 
log-periodic function of $a_{j}$ for negative $a_{j}$.

\section{STM equation}
\label{sec:STMeq}

The Skorniakov--Ter-Martirosian (STM) equation \cite{STM57} is an
integral equation that describes three-atom scattering interacting
through zero-range interactions. In this section, we discuss
the STM equation and relate the T-matrix element ${\cal T}^{\rm
  AD}_{ij}$ to the amplitude ${\cal A}_{ij}$ that is the solution to
the STM equation.

We consider only the S-wave
contribution and assume that higher partial wave contributions are
suppressed at low temperature.  For the three-fermion system, the STM
equation forms 9 coupled equations for the amplitudes ${\cal A}_{ij}
(k,p;E)$~\cite{Braaten:2008wd,Braaten:2009ey}.  For non-zero energy, the
equation is given by
\begin{eqnarray}
{\cal A}_{ij}(k,p;E) &=& (1 - \delta_{ij})\,Q(k,p;E)
\nonumber
\\
&&
+ \frac{2}{\pi} \sum_k (1 - \delta_{kj}) 
\int_0^\Lambda \! dq q^2\, Q(q,p;E) \, D_k(q;E) \,{\cal A}_{ik}(k,q;E) ,
\label{STMeq}
\end{eqnarray}
where ${\cal A}_{ij}(k,p;E)$ is the amplitude for an atom $i$ 
and a complementary pair of atoms to scatter into an atom $j$ and
a complementary pair and $\Lambda$ is an ultraviolet cutoff. 
The function $Q(k,p;E)$ 
and the 2-atom propagator $D_k(q;E)$ are given by
\begin{eqnarray}
Q(k,p;E)&=&\frac{1}{2kp}\ln\left[
\frac{k^2+kp+p^2-mE-i\epsilon} 
     {k^2-kp+p^2-mE-i\epsilon}
\right],
\label{func:Q}
\\
D_k(q;E)&=&\frac{1}{-1/a_k+\sqrt{\tfrac{3}{4}q^2-mE-i\epsilon}}.
\label{func:D}
\end{eqnarray}

The solutions of the STM equation \eqref{STMeq} depend
log-periodically on $\Lambda$ with a discrete scaling factor
$e^{\pi/s_0} \approx 22.7$, where $s_0 \approx 1.00624$. The dependence on
the arbitrary cutoff $\Lambda$ can be eliminated in favor of a
physical 3-body parameter such as the binding wave number of an Efimov
trimer in the unitary limit.  For convenience, we choose to work
directly with the wave number cutoff $\Lambda$ in our
calculations. If deep dimers are present,
the trimer has a finite width that allows it to decay
into an atom and a deep dimer.  The effects of deep dimers can be
taken into account by analytically continuing the cutoff $\Lambda$
into the complex plane \cite{Braaten:2002sr}
\begin{equation}                               
\Lambda\to \Lambda e^{i \eta_*/s_0},
\label{cutoff}
\end{equation}
where $\eta_*$ is a width parameter associated with the 
effects of the deep dimers.
$\Lambda$ and $\eta_*$ are determined
from experimental measurements of three-body recombination in the
$^6$Li system~\cite{Williams:2009zzd} and their numerical values are
$\Lambda=456~a_0^{-1}$ and $\eta_*=0.016$~\cite{Braaten:2009ey} 
where $a_0$ is the Bohr radius.

In order to obtain the T-matrix element, 
the amplitude ${\cal A}_{ij}$ must be multiplied with
the dimer-wavefunction renormalization factors $\sqrt{Z_i Z_j}$, 
where $Z_i=8\pi/a_i$,
\begin{equation}                               
{\cal T}^{\rm AD}_{ij}(k,p;E) 
=\frac{8\pi}{\sqrt{a_i a_j}}{\cal A}_{ij}(k,p;E).
\label{Tmatrix}
\end{equation}
By solving the STM equation \eqref{STMeq} and 
using the relation in Eq.~\eqref{Tmatrix}, 
we can calculate the relaxation rates
in Eqs.~\eqref{beta:opt} and \eqref{beta:sh}.

\section{Zero Temperature Results} 
\label{sec:results}
In this section we present our numerical results for the dimer
relaxation rate constants $\beta_i$ at zero temperature and compare
them with recent measurements in
Refs.~\cite{Lompe:2010,Nakajima:2010}.  We solve the STM equation in
Eq.~\eqref{STMeq} numerically with 5 input parameters: the three pair
scattering lengths $a_i$, $i=1,2,3$, the cutoff $\Lambda$ and the
width parameter $\eta_*$.  We use the values $\Lambda=456~a_0^{-1}$
and $\eta_*=0.016$ that have been determined from measurements of
the three-body recombination rate in Ref.~\cite{Williams:2009zzd} and have been
used in Ref.~\cite{Braaten:2009ey} to predict the atom-dimer
relaxation rate.

\begin{figure}[t]
\centerline{\includegraphics*[width=15cm,angle=0,clip=true]{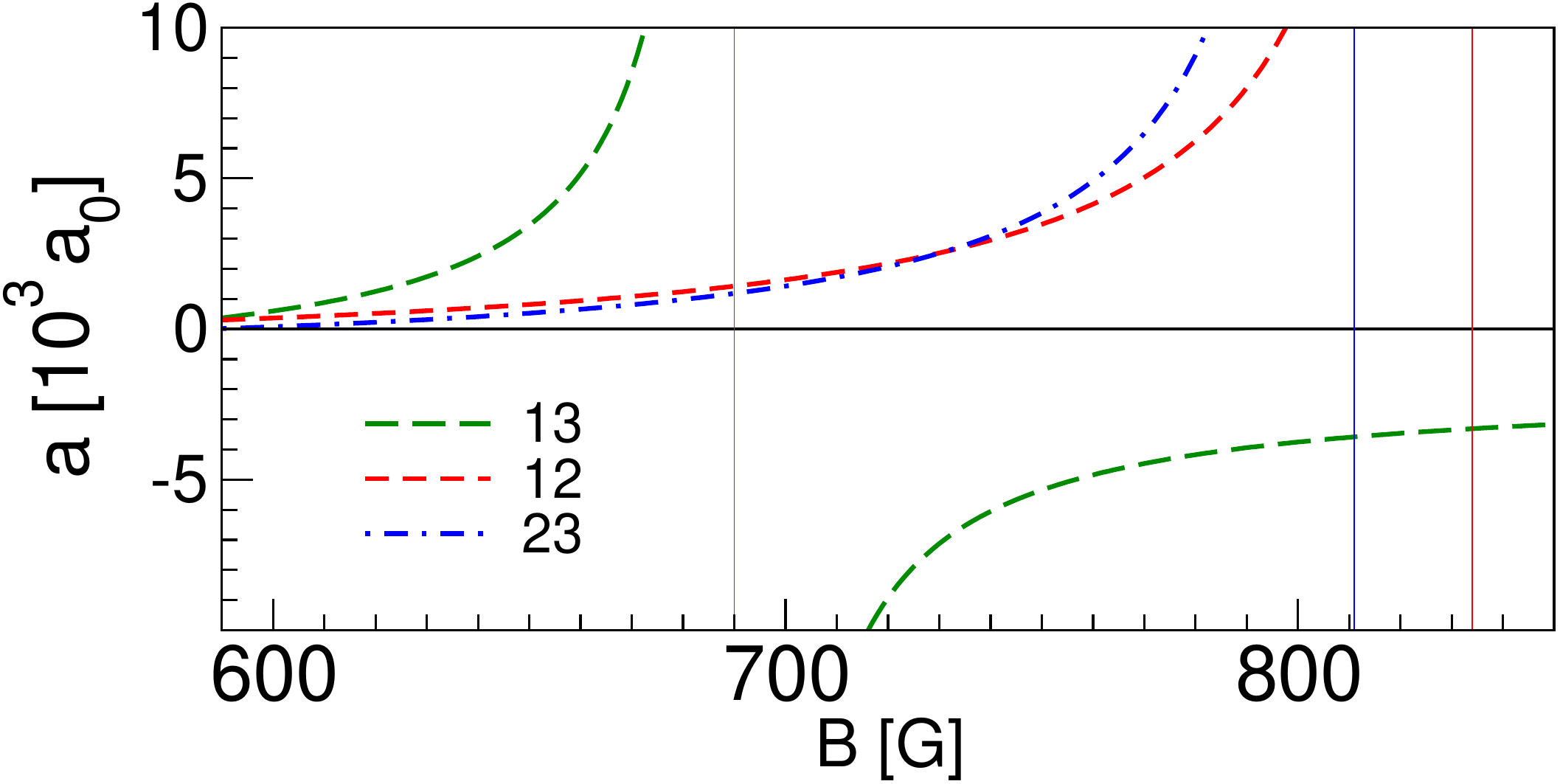}}
\vspace*{0.0cm}
\caption{(Color online) 
The scattering lengths for the three lowest hyperfine states of $^6$Li
in units of $10^3~a_0$ as a function of the magnetic field B 
from \cite{Julienne}:
$a_{13} \equiv a_2$ (long-dashed), $a_{12} \equiv a_3$ (short-dashed),
and $a_{23}  \equiv a_1$ (dashed-dotted).
The vertical lines at 690~G, 811~G, and 834~G
indicate the Feshbach resonances for $a_{13}$, $a_{23}$, and $a_{12}$,
respectively.}
\label{fig:aij}
\end{figure}
Fig.~\ref{fig:aij} shows the scattering lengths for the 3 lowest
hyperfine state of $^6$Li as a function of the magnetic field.  The
atom-dimer relaxation rate $\beta_i$ is non-zero in the region of
positive scattering length $a_i$.  The upper limit of the region is
set by the Feshbach resonances at 811~G, 690~G, and 834~G for $a_1$,
$a_2$, and $a_3$, respectively.  If the scattering length is much
larger than the van der Waals length $\ell_{\rm vdW}\approx 65 a_0$,
the universal theory is valid. We denote this region as the {\it
  universal region}. Corrections due to the finite range of the
interaction should be small in the universal region.  In practice, we
apply the universal zero-range theory 
when all scattering lengths are at least
two times larger than $\ell_{\rm vdW}$ corresponding to magnetic
fields $B>608$~G. The expected error due to finite range corrections
is $\ell_{\rm vdW}/a$ and therefore smaller than $50 \%$ in this region.

In Ref.~\cite{Braaten:2009ey}, two crossings of Efimov trimers with
the atom-dimer threshold have been predicted.  Both are located at the
1(23)-threshold: at $B_*=672$~G and at $B_*'\approx 597$~G. Here the
index 1 denotes the atom and the index (23) the dimer. Since the
atom-dimer relaxation is resonant when the trimer appears near the
threshold, two resonances are expected in the rate $\beta_1$ near
$B_*$ and $B_*'$.  The resonance at $B_*$ is well in the universal
region where all scattering lengths are much larger than $\ell_{\rm
  vdW}$ while $B_*'$ is slightly outside.  Therefore, the resonance
position $B_*$ should be accurately determined with corrections of
order $\ell_{\rm vdW}/a_{23}\approx 10$\% where the value of $a_{23}$
at the resonance was used.  The position $B_*'$ is outside the
universal region and can receive large non-universal corrections of
order 100\%. These error estimates are accurate up to a prefactor 
of order one. The exact value of this prefactor can only be obtained from
an explicit calculation of the range corrections. Note also 
that these percentage errors apply to the positions in terms of the 
scattering length. To obtain the errors for the corresponding magnetic 
field they have to be converted using Fig.~\ref{fig:aij}.

\begin{figure}[t]
\centerline{\includegraphics*[width=15cm,angle=0,clip=true]{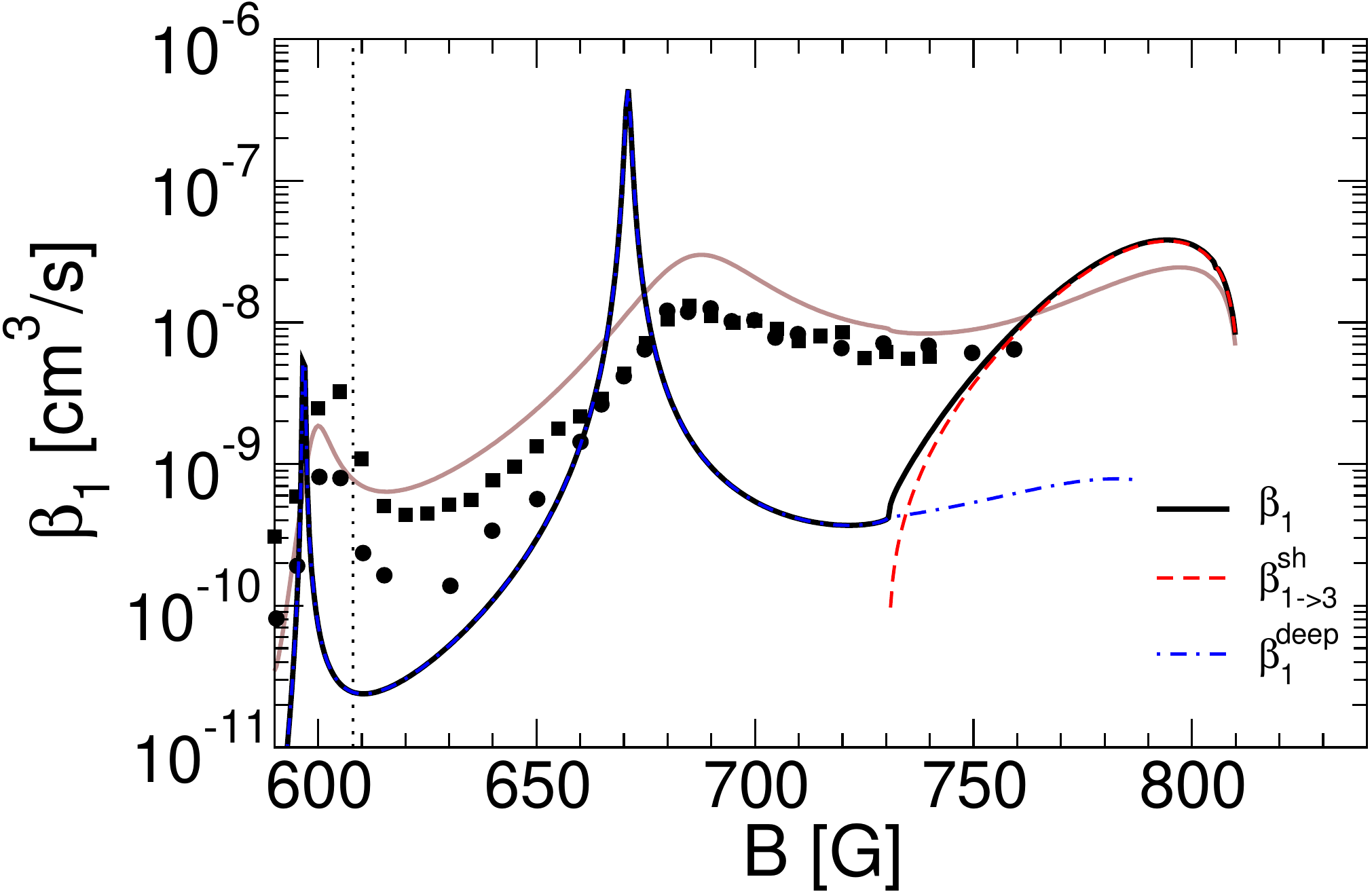}}
\vspace*{0.0cm}
\caption{(Color online) 
The relaxation rate constant for the 23-dimer and atom 1
as a function of the magnetic field $B$.
The squares and circles are data points from Ref.~\cite{Lompe:2010} and 
Ref.~\cite{Nakajima:2010}, respectively. 
The curves are our results for the total rate $\beta_1$ (solid line),
the partial rate into atom 3 and the 12-dimer $\beta^{\rm sh}_{1\to3}$ 
(dashed line),  and the rate into an atom and a deep dimer 
$\beta^{\rm deep}_1$(dashed-dotted line). 
The light solid line gives the  total rate $\beta_1$ for the 
parameters obtained in Ref.~\cite{Lompe:2010} while the
vertical line marks the boundary of the universal region.
}
\label{fig:beta1}
\end{figure}
In Fig.~\ref{fig:beta1}, we show our numerical results for $\beta_1$ and
compare them with the recent measurements of
Refs.~\cite{Lompe:2010,Nakajima:2010}. We give the full relaxation rate as
well the individual contributions from shallow and deep dimers. In the
magnetic field region from $590$~G to $730$~G, only the relaxation into deep
dimers contributes to $\beta_1$ since the energy of the $23$-dimer is larger
than the energy of the other shallow dimers. At a magnetic field of $730$~G,
the $23$-dimer crosses the $12$-dimer and the relaxation channel into
$12$-dimers opens up. As the crossing is approached from above,
the rate vanishes as $\sqrt{a_1-a_3}$ in agreement with the analytical
result from Sec.~\ref{sec:relaxation}.
Between 790 and 810~G, the condition $a_1 \gg a_3$ with $a_2$ non-resonant
is satisfied approximately. The rate into the shallow dimer in this region
scales approximately as $a_3^2/a_1$ in agreement with the prediction
of D'Incao and Esry \cite{DIncao:2009}.
At $811$~G, the $23$-dimer disappears through the
three-atom threshold and the relaxation rate vanishes.

Our results show two resonances at $B_*=672$~G and $B_*'\approx
597$~G.  There is also a dramatic change in the relaxation rate at
$730$~G because the relaxation channel into $12$-dimers opens and the
corresponding rate into the $12$-dimer increases rapidly.  Our results
describe the resonances in the experimental data qualitatively.  The
second resonance has been measured at $B_{*}^{\rm exp}=685$~G in
Ref.~\cite{Lompe:2010,Nakajima:2010}. This value is 13~G away from the
theoretical prediction $B_*=672$~G. In terms of the scattering length,
this corresponds to
$a_{23}(B_{*}^{\rm exp})/a_{23}(B_{*})=1076/835\approx 1.3$,
leading
to a 30\% shift in the resonance position.  This shift is a factor
three larger than the naive error estimate of 10\%.  Taking into
account the unknown prefactor of order one in the estimate, however,
the two values are consistent. Except near the resonance, the
experimental data are generally above our results.
 
Lompe {\it et al.}~\cite{Lompe:2010} analyzed their data near the resonance at
$B_*$ using an approximate analytic expression from~\cite{Braaten:2009ey} and
extracted the resonance position and the width parameter $\eta_*$. They found
the value $\eta_*=0.34$ which is more than an order of magnitude larger than
the value 0.016 extracted from three-body recombination. Moreover, the
normalization of the relaxation rate was adjusted to describe the data.  The
light solid curve in Fig.~\ref{fig:beta1} gives our universal result for the
parameters $\Lambda=329~a_0^{-1}$ (which reproduces the resonance position
$B_{*}^{\rm exp}=685$~G) and $\eta_*=0.34$. These parameters give a much
better description of the data in Refs.~\cite{Lompe:2010,Nakajima:2010} but
are generally a factor 2-3 above the experimental data. However, one should
keep in mind that the experimental data are only a factor three lower than the
unitarity bound and finite temperature effects are likely important. We will
come back to this issue in the next section.

The reason for the considerably larger value of $\eta_*$ extracted in
\cite{Lompe:2010} compared to the value from recombination data is not
understood. However, we note that a similar discrepancy between the values of
$\eta_*$ from atom-dimer relaxation and three-body recombination occurs in the
bosonic system of $^{133}$Cs atoms \cite{Knoop:2008,Helfrich:2009uy}.

Nakajima {\it et al.}\ \cite{Nakajima:2010} performed a 
numerical analysis for $\beta_1$ based on the universal theory and on a
two-channel model.  Their results obtained with the universal theory agree
with our calculation.  Within the two-channel model they derived
energy-dependent scattering lengths which introduce non-universal effects in
the two-body amplitudes.  Because this model cannot resolve the discrepancy
between the universal results and the measurements they concluded that the
three-body parameters $\Lambda$ and $\eta_*$ depend on the magnetic field.

\begin{figure}[t]
\centerline{\includegraphics*[width=15cm,angle=0,clip=true]{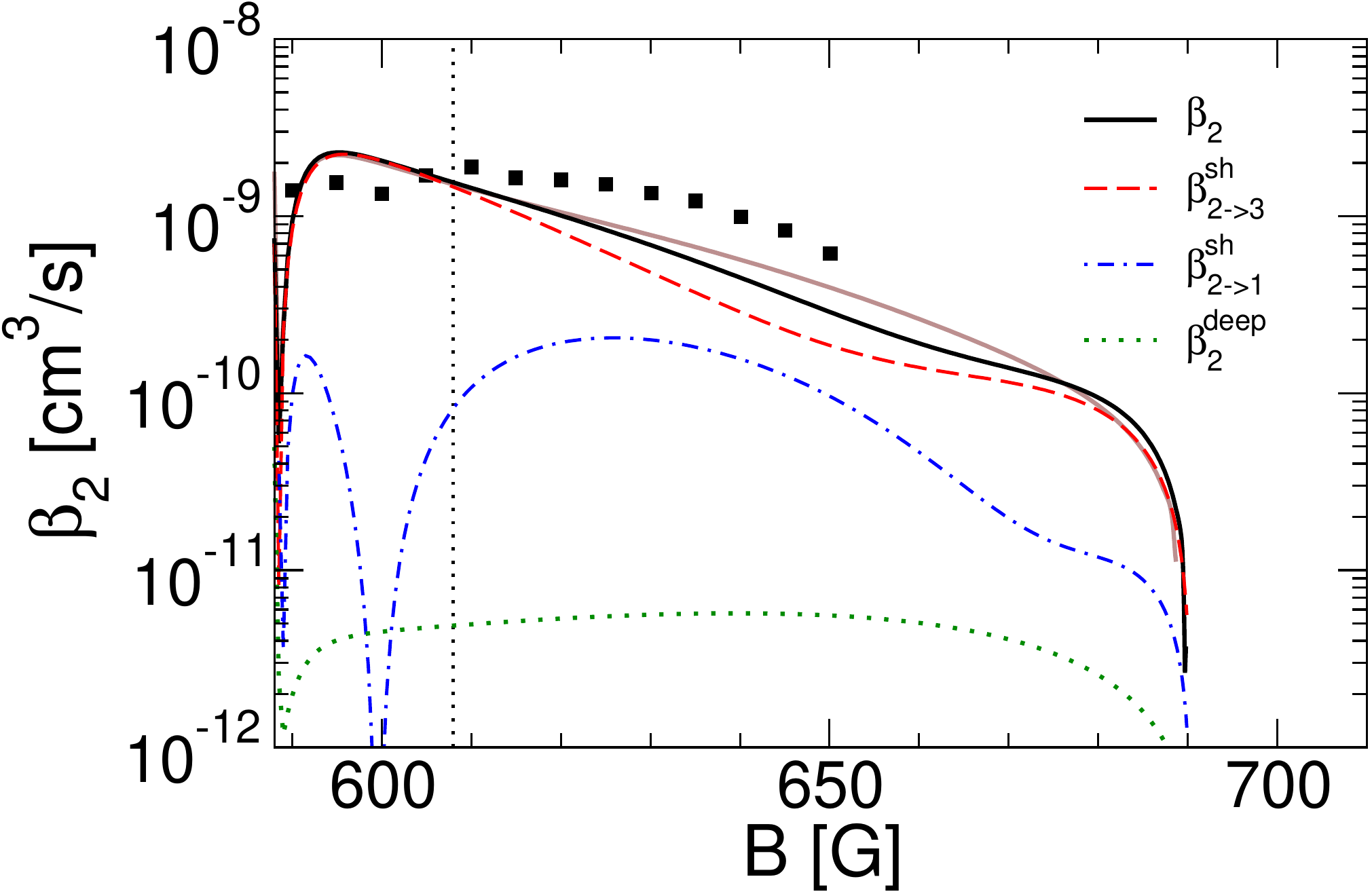}}
\vspace*{0.0cm}
\caption{(Color online) 
The relaxation rate constant for the 13-dimer and atom 2
as a function of the magnetic field $B$. 
The squares are data points from Ref.~\cite{Lompe:2010}.
The curves are our results for the total rate $\beta_2$ (solid line)
and the partial rates into an atom 3 and a 12-dimer $\beta^{\rm sh}_{2\to 3}$ 
(dashed line), into an atom 1 and a 23-dimer $\beta^{\rm sh}_{2\to 1}$ 
(dashed-dotted line), and into an atom and a deep dimer 
$\beta^{\rm deep}_2$ (dotted line).
The light solid line gives the  total rate $\beta_1$ for the 
parameters obtained in Ref.~\cite{Lompe:2010} while the
vertical line marks the boundary of the universal region.
}
\label{fig:beta2}
\end{figure}
In Fig.~\ref{fig:beta2}, we show the experimental data from
Ref.~\cite{Lompe:2010} and our numerical results for $\beta_2$. The
total rates, both of the data and the numerical results, show no
pronounced structure.  However, there is a local minimum in the
partial rate $\beta^{\rm sh}_{2\to 1}$ near $600$~G that is outside
the universal region.  As discussed in Ref.~\cite{DIncao:2009}, this
minimum is the effect of destructive interference between 
different recombination channels. This
interference pattern does not exist in
systems with identical bosons where relaxation can occur only into
deep dimers. In the total rate, the interference is hidden by the
dominant process $\beta^{\rm sh}_{2\to 3}$.
The light solid curve again gives our
universal result for the parameters $\Lambda=329~a_0^{-1}$ 
and $\eta_*=0.34$ obtained in Ref.~\cite{Lompe:2010}.
The difference between the two parameter sets is very small 
for $\beta_2$, but the alternative set gives a slightly
better description of the data. 
Between 670 and 690~G, $a_2$ is much larger than
$a_1$ and $a_3$ which are approximately equal. 
The scaling of the calculated rates into the shallow dimers 
is consistent with the prediction 
of D'Incao and Esry \cite{DIncao:2009}:
$\beta^{\rm sh}_{2\to 3}\sim a_3^2/a_2$ and 
$\beta^{\rm sh}_{1\to 3}\sim a_1^2/a_2$.

\begin{figure}[t]
\centerline{\includegraphics*[width=15cm,angle=0,clip=true]{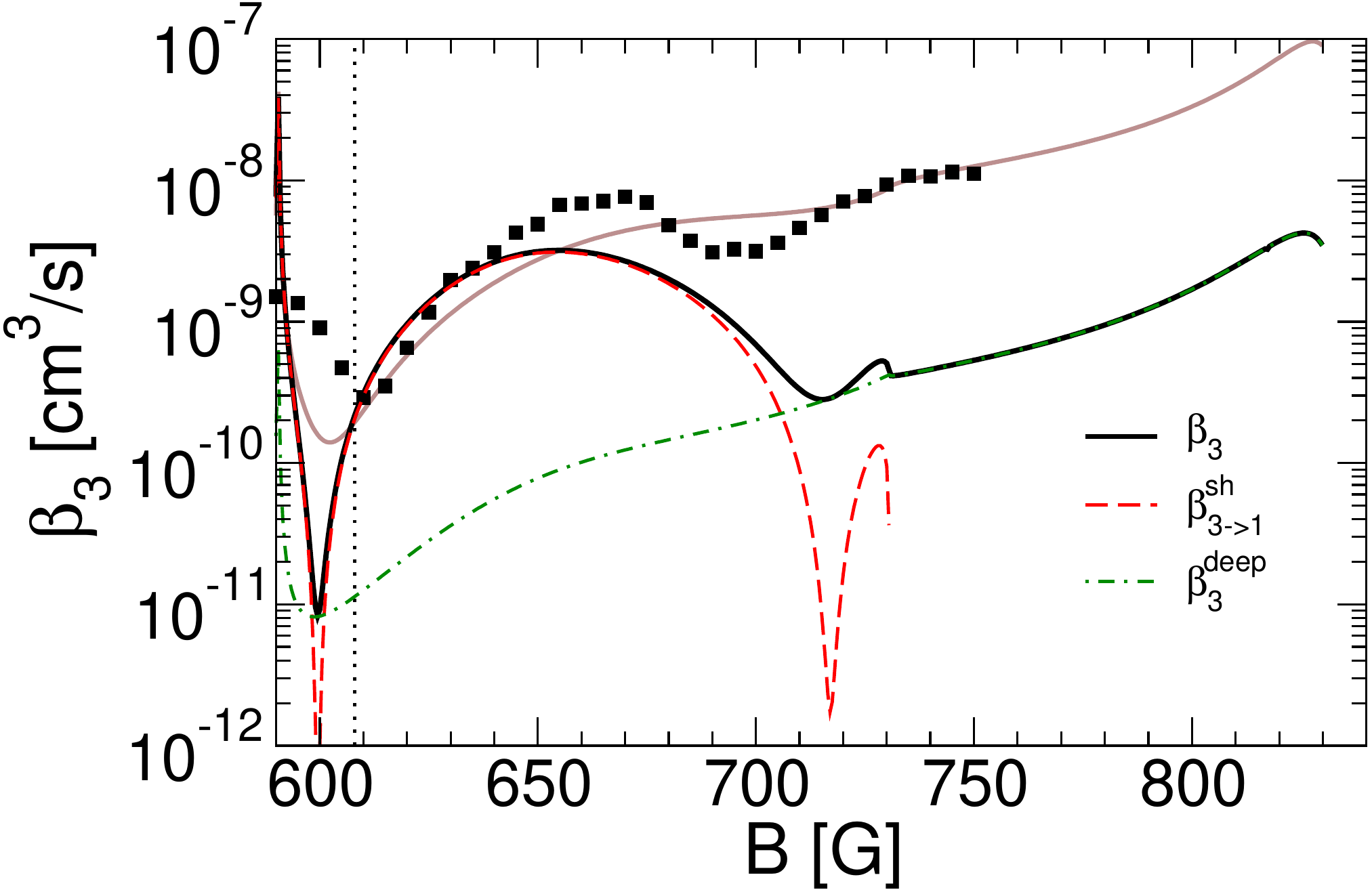}}
\vspace*{0.0cm}
\caption{(Color online) 
The relaxation rate constant for the 12-dimer and atom 3 as a function of 
the magnetic field $B$.
The squares are data points from Ref.~\cite{Lompe:2010}.
The curves are our results for the total rate $\beta_3$ (solid line),
the partial rate into atom 1 and a 23-dimer $\beta^{\rm sh}_{3\to 1}$ 
(dashed line), and the rate into an atom and a deep dimer $\beta^{\rm deep}_3$ 
(dashed-dotted line).
The light solid line gives the  total rate $\beta_1$ for the 
parameters obtained in Ref.~\cite{Lompe:2010} while the
vertical line marks the boundary of the universal region.
}
\label{fig:beta3}
\end{figure}
Fig.~\ref{fig:beta3} shows our results and the recent measurement
\cite{Lompe:2010} for $\beta_3$.  Between $590$~G and $730$~G,
the energy of the $12$-dimer is smaller than the energy of the $23$-dimer
and the relaxation channel into $23$-dimers is open.  After the $23$-dimer
crosses the $12$-dimer at $730$~G this relaxation channel is closed and
only relaxation into deep dimers is possible. 
As the crossing is approached,
the rate vanishes as $\sqrt{a_3-a_1}$ in agreement with the analytical
result from Sec.~\ref{sec:relaxation}. 
Between 832 and 834~G, $a_1$ is negative and
the condition $a_3 \gg |a_1|$ with $a_2$ non-resonant
is satisified approximately. The rate into deep dimers
scales in this region approximately as $a_1^2/a_3$ consistent with
the prediction by D'Incao and Esry.
Two
interference minima have been observed at $610$~G and $695$~G
\cite{Lompe:2010} while our results show two minima at $600$~G and
$715$~G.  Above $700$~G, the data are larger than our results 
by more than a factor 10.
Using the alternative parameters $\Lambda=329~a_0^{-1}$
and $\eta_*=0.34$ obtained in Ref.~\cite{Lompe:2010},
we again find a better agreement with the data.
With these parameters the second minimum in the rate into the shallow dimer
disappears beyond 730~G. Hence, the partial rate decreases monotonically
and vanishes near 730~G. For smaller $\eta_*$, the position of the minimum
in the total rate is around 730~G. As $\eta_*$ increases, the position
remains almost the same and the depth of the minimum becomes shallower.
The minimum is not visible when $\eta_*=0.34$ because the rate into
deep dimers is much larger than the rate into the shallow dimer.
Therefore, with these parameters that were fit to data for $\beta_1$
the position of the second minimum in the data for $\beta_3$ cannot be explained
correctly.

If the rate into 23-dimers could be separated experimentally from the
total rate, it would clearly determine the positions of the local
minima. This could be achieved  
by tuning the depth of trapping potential such that it
is much larger than the kinetic energies of atoms
and 23-dimers in the final state but much smaller than the energies of
atoms and deep dimers in the final state.  The kinetic energies of an atom
and a deep dimer in the final state could be estimated from the
binding energy of the deep dimers.  Their energies would be of the order of
the van der Waals energy or larger: $E_{\rm vdW}/h
\approx 154$~MHz, where 
$E_{\rm vdW}=\hbar^2/(m\ell_{\rm vdW}^2)$.\footnote{
A convenient conversion constant for $^6$Li atoms is given by
$\hbar^2/(m a_0^2) = 600h$~GHz= 28.8$k_B$~{\rm K}. }
The kinetic energies of an atom and a 23-dimer are given by the difference in
binding energies between the 12-dimer and the 23-dimer:
$\hbar/(2\pi m)(a_{23}^{-2}-a_{12}^{-2})$ is about $1$~MHz at 650~G 
and vanishes
at 730~G.  This way one may be able to measure the rate into
deep dimers separately and to extract the rate into the 23-dimer. 

\section{Finite Temperature Results}
\label{sec:themal}
The results from the previous section suggest that finite temperature
effects may play an important role in understanding the
atom-dimer relaxation data from Refs.~\cite{Lompe:2010,Nakajima:2010}.
A full finite temperature calculation of the $^6$Li system is beyond 
the scope of this work. Therefore, we perform an approximate 
calculation of the relaxation rate $\beta_1$ near the resonance
at $B_{*}^{\rm exp}=685$~G.

\begin{figure}[t]
\centerline{\includegraphics*[width=15cm,angle=0,clip=true]{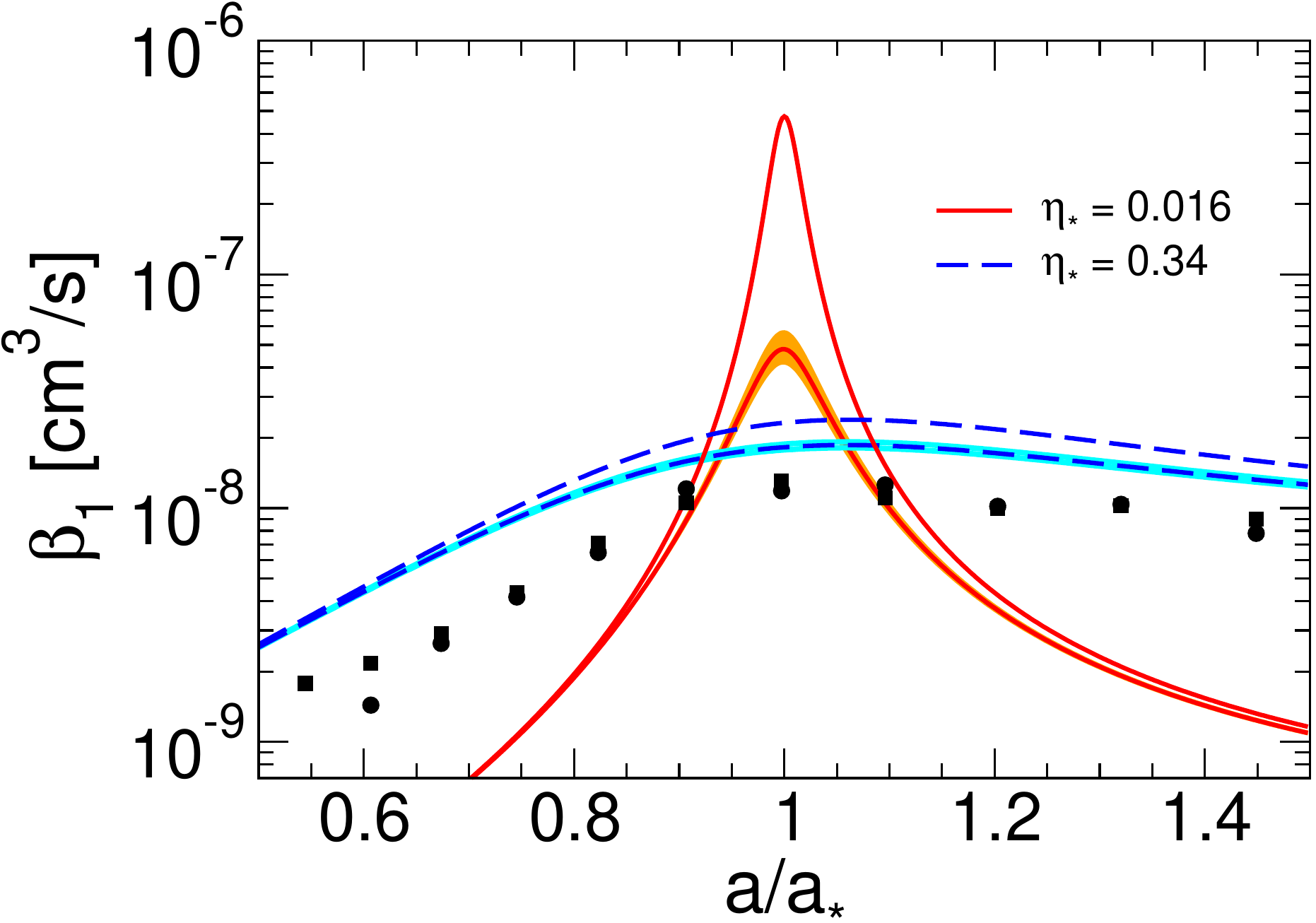}}
\vspace*{0.0cm}
\caption{(Color online) 
The relaxation rate constant $\beta_1$ for a 23-dimer and an atom 1
as a function of $a/a_*$ near the resonance at $B_{*}^{\rm exp}=685$~G
($a_*=1076~a_0$).
Squares and circles are data points from Ref.~\cite{Lompe:2010} and 
Ref.~\cite{Nakajima:2010}, respectively. 
The solid (dashed) curves correspond to $\eta_*=0.016$ ($\eta_*=0.34$). 
The upper curves give the zero temperature result, while the lower curves 
give the finite temperature result for $T=60 \pm 15$ nK with the shaded
area indicating the temperature uncertainty.
}
\label{fig:beta1_T}
\end{figure}

We start from the approximate expression for the scattering length
between an atom 1 and a 23-dimer near the resonance at $B_{*}$ that
was extracted from a calculation of the trimer binding energy in
\cite{Braaten:2009ey}:
\begin{equation}
a_{1(23)} \approx \left( C_1 \cot[ s_0 \ln(a_{23}/a_*) + i \eta_*]
                         + C_2 \right) a_{23} ,
\label{aAD-a23}
\end{equation}
where $a_* = a_{23}(B_*)$ and the coefficients are 
$C_1 = 0.67$ and $C_2 = 0.65$.
Using the scattering length approximation for the S-wave atom-dimer 
scattering amplitude in the 1(23)-channel,
\begin{equation}
f_{1(23)}(k) = \left[ -1/a_{1(23)}-ik \right]^{-1},
\label{fAD-a23}
\end{equation}
we can calculate the inelastic scattering cross section. At low
temperatures, the contributions from higher partial waves can be
neglected.  We subtract the elastic cross section from the total cross
section obtained via the optical theorem as in \cite{Braaten:2006nn}
and find for the inelastic cross section:
\begin{equation}
\sigma_{1(23)}^{(inelastic)}(k) = 
\frac{4\pi}{k} \frac{-{\rm Im}\, a_{1(23)}}{1-2k{\rm Im}\, a_{1(23)}
+k^2 |a_{1(23)}|^2}.
\label{sigAD-a23}
\end{equation}
The total dimer-relaxation rate $\beta_1$ can then be calculated
by taking a Boltzmann thermal average of the inelastic reaction rate
$v_{rel}\, \sigma_{1(23)}^{(inelastic)}(k)$
where $v_{rel}=3\hbar k/(2m)$ is the relative velocity of the atom and
dimer in the initial state. This leads to the expression
\begin{equation}
\beta_1(T)=\frac{3\hbar}{2m}\langle k\sigma_{1(23)}^{(inelastic)}(k) \rangle=
\frac{3\hbar}{2m} \frac{3\lambda_T^3}{4\pi^2}\sqrt{\frac{3}{2}}
\int_0^\infty k^2 dk \;k\sigma_{1(23)}^{(inelastic)}(k)\;
e^{-\frac{3\hbar^2k^2}{4m k_BT}},
\label{betaAD-a23}
\end{equation}
where $\lambda_T=\sqrt{2\pi \hbar^2/(m k_B T)}$ is the thermal
de Broglie wavelength of the atoms.

Our results for the atom-dimer relaxation rate constant near the
resonance at $B_{*}^{\rm exp}=685$~G are shown in
Fig.~\ref{fig:beta1_T}.  The results for the width parameters
$\eta_*=0.016$ and $\eta_*=0.34$ are given by the solid and dashed
curves, respectively.  In each case the upper curves give the zero
temperature result while the lower curves give the finite temperature
result for $T=60 \pm 15$ nK. Here, the shaded area indicates the
uncertainty from the temperature.  The value of $a_*=1076~a_0$ has
been fixed to reproduce the resonance position at $B_{*}^{\rm
  exp}=685$~G.  For $\eta_*=0.016$, the finite temperature effects
decrease the height of the peak by an order of magnitude but the
predicted resonance shape is still much narrower than the data. The
finite temperature effects are much less severe for
$\eta_*=0.34$. They only lead to a reduction of $\beta_1$ by about
25 \% but clearly improve the description of the data. We
conclude that finite temperature effects can not
resolve the question of the different values for $\eta_*$ in the
three-body recombination and dimer-relaxation data. Moreover, while
finite temperature effects are important, a qualitative description of
the data at $T\approx 60$ nK can be already achieved with a zero
temperature calculation.

\section{Summary and Outlook}
\label{sec:summary}
In this work, we have studied Efimov physics in
atom-dimer relaxation of $^6$Li atoms in the three lowest hyperfine
states using the universal zero-range theory. 
Two resonances were observed at magnetic fields 603~G and
685~G in the relaxation rate $\beta_1$ in recent experiments
\cite{Lompe:2010,Nakajima:2010}. These resonances are consequences of
two Efimov trimers close to the atom-dimer threshold. Their positions
have been predicted by Braaten {\it et al.}\
\cite{Braaten:2009ey}. The measured position of the resonance at 685~G
\cite{Lompe:2010,Nakajima:2010}, which is well within the universal
region, is larger than the prediction by about 30\%.  This is
consistent with an error of order 10\% error due to effective range
corrections. However, the value $\eta_*=0.016$ extracted from the
three-body recombination data is not able to describe the atom-dimer
relaxation data which require the larger value $\eta_*=0.34$
\cite{Lompe:2010}. The reason for the larger value of $\eta_*$ in
dimer relaxation compared to the value from recombination data is not
understood.  However, we note that a similar discrepancy between the
values of $\eta_*$ from atom-dimer relaxation and three-body
recombination occurs in the bosonic systems of $^{133}$Cs atoms
\cite{Knoop:2008,Helfrich:2009uy}.

Using the value $\eta_*=0.34$, our zero temperature calculation is able to
describe the data qualitatively.  In the vicinity of this resonance, we have
also performed an approximate finite temperature calculation and find sizable
temperature effects that can suppress the relaxation rate by an order of
magnitude if $\eta_*=0.016$. For the larger value $\eta_*=0.34$, however,
these effects lead to a moderate suppression of about 25\%, such
that zero temperature results are useful as a first approximation.

In Ref.~\cite{Lompe:2010}, also two local minima at 610~G and 695~G are
discovered in the rate $\beta_3$. Those minima can be associated with
destructive interference between different recombination
channels~\cite{DIncao:2009}.  Our numerical results show that the partial rate
into 23-dimers is responsible for the minima but the positions of the minima
are displaced by -10~G and +20~G from the measurements. These displacements
correspond to a 30\% shift in terms of the scattering length 
compared to our predictions. Since the scattering length, $a_{23}$, is 
about a factor two larger in this magnetic field region
than around the resonance in $\beta_1$, one would expect smaller
corrections here. The observed shifts
indicate that the destructive interference leading to the minima 
might be dominated by wave numbers 
$k$ larger than $1/a_{23}$ such that corrections of 
order $k\ell_{vdW}$ are important. Moreover, finite temperature effects
could fill the minima in an asymmetric fashion.
%
%
The total rate $\beta_2$ shows no structure.
However, the partial rate into the 23-dimer shows a local minimum near 600~G.
In the total rate this feature is hidden by the dominant rate into the
$12$-dimer.

In order to better understand the discrepancy between the values of
$\eta_*$ and the resonance positions extracted from three-body
recombination and atom-dimer relaxation two important improvements are
required in a future analysis. First, a full finite temperature
calculation of atom-dimer relaxation should be carried out. This
requires calculating atom-dimer scattering above the dimer breakup
thresholds and in higher partial waves.  Due to the different pair
scattering lengths in the three channels, such a calculation is
considerably more complex than in the case of identical bosons
\cite{Braaten:2008kx}.  However, it will allow to better distinguish
effects from the resonance width parameter $\eta_*$ and from the
finite temperature which can be partially traded for each other
\cite{Braaten:2008kx,Helfrich:2009uy}.  Second, 
an analysis of the effective range
corrections should be performed in order to describe the observed
shifts in the resonance positions quantitatively. A similar analysis
for systems of identical bosons was carried out in
Refs.~\cite{Hammer:2006zs,Ji:2010su}.

\begin{acknowledgments}
We thank Eric Braaten, Selim Jochim, and Thomas Lompe for discussions and
Selim Jochim for providing their experimental results.
We acknowledge the INT
  program ``Simulations and Symmetries: Cold Atoms, QCD, and
  Few-hadron Systems'', during which this work was carried out.
This work was supported in part by a joint grant from AFOSR and ARO,
by the BMBF under contract 06BN9006 and by DOE grant DE-FG02-00ER41132.
\end{acknowledgments}

\end{document}